# Mid-infrared Fourier transform spectroscopy with a broadband frequency comb

Florian Adler,[1,*] Piotr Masłowski,[1,3] Aleksandra Foltynowicz,[1] Kevin C. Cossel,[1] Travis C. Briles,[1] Ingmar Hartl,[2] and Jun Ye[1]

[1]*JILA, National Institute of Standards and Technology and University of Colorado, and Department of Physics, University of Colorado, Boulder, Colorado 80309-0440, USA*
[2]*IMRA America Inc., 1044 Woodridge Avenue, Ann Arbor, Michigan 48105, USA*
[3]*Permanent address: Instytut Fizyki, Uniwersytet Mikołaja Kopernika, ul. Gagarina 11, 87-100 Toruń, Poland*
[*]*fadler@jila.colorado.edu*

**Abstract:** We present a first implementation of optical-frequency-comb-based rapid trace gas detection in the molecular fingerprint region in the mid-infrared. Near-real-time acquisition of broadband absorption spectra with 0.0056 cm$^{-1}$ maximum resolution is demonstrated using a frequency comb Fourier transform spectrometer which operates in the 2100-to-3700-cm$^{-1}$ spectral region. We achieve part-per-billion detection limits in 30 seconds of integration time for several important molecules including methane, ethane, isoprene, and nitrous oxide. Our system enables precise concentration measurements even in gas mixtures that exhibit continuous absorption bands, and it allows detection of molecules at levels below the noise floor via simultaneous analysis of multiple spectral features.

## References and links

## 1. Introduction

Optical spectroscopy has been an indispensible tool for molecular analysis in physics, chemistry, and other disciplines in the last decades, but modern applications in fields such as atmospheric science and medical analysis are calling for increasingly powerful experimental techniques. For example, rapid and simultaneous monitoring of trace concentrations and chemical reaction dynamics of atmospheric gasses could improve climate models for predicting the effects of global warming [1-4]. The diagnosis of diseases via biomarkers in human breath will require a method that is capable of detecting many different molecules in real-time and in a wide range of concentrations [5,6]. These examples push the limits of conventional Fourier transform infrared [7] (FTIR) and continuous-wave (cw) laser spectrometers [8]. Consequently, there is a strong interest in more advanced spectroscopic instruments that are able to combine high detection sensitivity, broad detection bandwidth, access to a wide variety of molecular species, high resolution, and short acquisition times [9-12]. In particular, combining the latter two requirements represents the outstanding challenge for standard FTIR due to its typical incoherent light source.

Fourier transform spectrometers based on optical frequency combs (FC-FTS), which offer excellent spectral brightness and spatial coherence, have been introduced to address these problems [13-21]. In addition to using FC-FTS based on a Michelson interferometer with a mechanical stage [15-17,19,20], dual-comb spectrometers have been implemented by using two femtosecond lasers with slightly different repetition rates, thus mimicking the effect of a fast-scanning delay stage [14,18,21,22]. Both approaches have been used for proof-of-principle studies in the near infrared [17-22] relying on weak vibrational overtone transitions of just a few molecular species (mostly, $C_2H_2$, $CO_2$, and HCN) that have sufficient line strengths to be detected at good signal-to-noise (S/N) ratios. For FC-FTS to truly replace FTIR, these new spectroscopic systems need to be extended into the important molecular fingerprint region in the mid-infrared between 1000 cm$^{-1}$ and 4000 cm$^{-1}$. The fundamental rovibrational transitions in this spectral range are much stronger than the near-infrared overtones; thus, a mid-IR comb spectrometer promises drastic improvements in detection limit and coverage of species; however, the lack of direct femtosecond laser sources in this region has complicated the development of such an instrument. Although mid-infrared comb spectroscopy has been demonstrated based on difference frequency generation [14] or optical parametric oscillators (OPOs) [15,16], these studies were not able to provide sufficient performance to compete with established techniques.

In this article we present an advanced FC-FTS based on a high power OPO that operates in the important mid-infrared window from 2100 cm$^{-1}$ to 3700 cm$^{-1}$. The system provides for the first time a combination of resolution, spectral range, sensitivity, and acquisition speed that is sufficient for detecting trace quantities of a wide range of molecules under real-world conditions. The experimental achievements include reliable concentration measurements of mixtures of multiple species with overlapping absorption bands or vastly different abundances, as well as molecules with irresolvable continuous absorption features. Our system is capable of achieving part-per-billion (ppb)-level sensitivity with an integration time of 30 s or less for a variety of important molecules, such as the greenhouse gases methane, carbon dioxide, and nitrous oxide [23-25], as well as isoprene and formaldehyde, which are important contributors to pollution [26,27]. Furthermore, we demonstrate detection of ethane and methanol, which are subjects of interest in human breath analysis [28-30]. These molecules represent just a small set of trace gasses within the detection range of this new system.

## 2. Experimental setup and procedure

Fig. 1 shows a schematic of the experimental setup. The high-power light source of our mid-infrared FC-FTS is a fiber-laser-pumped OPO frequency comb optimized for mid-infrared light emission. The center frequency of the idler comb is tunable from 2100 cm$^{-1}$ to 3600 cm$^{-1}$ (~2.8-4.8 μm) with a mode spacing of 136 MHz and a power per comb line on the order of 1-10 μW (>1 W of maximum average power). The simultaneous spectral coverage ranges from 50 cm$^{-1}$ at a centre frequency of 2300 cm$^{-1}$ to 400 cm$^{-1}$ at 3500 cm$^{-1}$. A detailed description of the OPO comb may be found in Ref. [31]. To minimize drifts of the spectrum during the measurement, the OPO comb is phase-stabilized to the pump laser, whose repetition rate is locked to a microwave reference for long-term stability. Using a phase-locked OPO comb also ensures low intensity noise compared to non-comb supercontinuum sources.

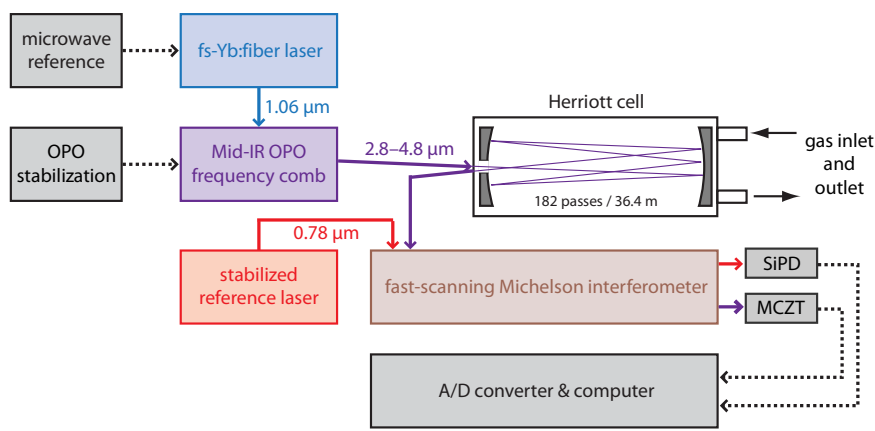

Fig. 1. Schematic of the experimental setup; details on the OPO stabilization can be found in Ref. [31]; SiPD, silicon photo diode; MCZT: mid-IR HgCdZnTe detector. Solid lines indicate optical paths, dotted lines represent electronic lines.

The light from the OPO is coupled into a Herriott multi-pass cell (New Focus 5611) [32] with 0.3 l volume and a total path length of 36.4 m (182 passes), which contains the sample or reference gas, respectively. The spatial coherence of the laser source makes the coupling simple and efficient. The light exiting the multi-pass cell is directed to a home-built FTS which comprises a rapid-scanning Michelson interferometer with a thermoelectrically-cooled, photovoltaic MCZT (HgCdZnTe) detector (Vigo Systems PVI-2TE-5) [32]. The high power comb source eliminates the need for a highly sensitive liquid-nitrogen-cooled IR detector, simplifying daily use of the system. In addition, the collimated laser light provides an important advantage in étendue over a standard FTIR spectrometer. Alongside the mid-IR beam, a rubidium-stabilized diode laser with a wavelength of 780 nm is coupled into the interferometer for absolute frequency calibration of spectroscopic signals. The interferograms of both mid-IR light and 780-nm reference laser are recorded simultaneously with a 2-channel data acquisition card (National Instruments NI 5922) [32] at 1 MSamples/s and 22 bit resolution. The high data acquisition rate allows us to exploit the scanning speed of the FTS; to avoid digitization noise in the analog-to-digital conversion, at least 20 bit resolution are required. From the interferogram of the reference laser we determine the positions of the zero-crossings, which are used to resample the mid-infrared trace for exactly evenly spaced data points. We acquire double-sided interferograms to get phase-corrected Fourier spectra [7] after performing complex fast Fourier transform. To extract the absorption spectrum, we measure first the laser spectrum through the sample gas $I_1(\nu)$, then purge the cell, fill it back to the original pressure with $N_2$, and measure the reference spectrum $I_0(\nu)$. The spectral absorbance $\alpha(\nu)$ is then determined via $\alpha(\nu) = -L^{-1} \ln\left[I_1(\nu)/I_0(\nu)\right]$, where $L$ is the optical path length inside the multi-pass cell. The gas concentrations are extracted from the data via a numerical fit of a model spectrum calculated with spectral data (line positions, line strengths, pressure broadening and shift coefficients) from HITRAN [33] or reference spectra from NWIR [34] (when HITRAN data was not available) based on a modified Marquardt algorithm [35]. The parameters for the fitted Voigt profile are a concentration value for each sample species, a polynomial baseline, and a frequency correction (to account for misalignment of the beams inside the interferometer, typically on the order of $10^{-3}$ cm$^{-1}$). The uncertainty value for the concentration is extracted from the covariance matrix.

## 3. Results

### 3.1 Measurement of individual molecular species

To demonstrate the versatility and wide spectral coverage of our spectrometer we recorded spectra of a variety of molecular species (each diluted in pure $N_2$) in different spectral regions spanning from 2170 cm$^{-1}$ to 3520 cm$^{-1}$. Figure 2 shows measured spectra (black) of nitrous oxide [9.0 ppm in Fig. 2(a) and 142 ppm in Fig. 2(f)], 58 ppm formaldehyde [Fig. 2(b)], 11 ppm ethane [Fig. 2(c)], 10 ppm methane [Fig. 2(d)], and 16 ppm isoprene [Fig. 2(e)]. With the exception of the high-resolution spectrum in Fig. 2(d), all of these spectra were recorded with integration times of 30 s or less. Each plot also shows the fitted spectrum using data from HITRAN or NWIR, as indicated in each graph (plotted in purple and negative for clarity). These measurements show the instrument's ability to rapidly cover a wide spectral region with high resolution and extract concentrations regardless of the structure of the absorption spectrum. For example, isoprene exhibits a continuous absorption band, which makes quantitative measurements susceptible to errors from baseline drifts. Being able to determine the concentration of such a sample is extremely valuable since molecules with continuous bands (e.g., isoprene, methanol, ethanol, acetone, etc.) are highly important for many applications.

### 3.2 Instrument performance limits

We have carried out systematic characterizations of the limits of our system in terms of resolution, measurement accuracy, absorption sensitivity, and concentration detection limit. From measurements of the noise floor we obtain an absorption sensitivity of $3.8 \times 10^{-8}$ cm$^{-1}$ Hz$^{-1/2}$ per spectral element [19]. The number of elements is given by the optical bandwidth divided by the spectral resolution; for example, at a centre frequency of 3000 cm$^{-1}$, we detect ~45,000 simultaneous channels at maximum resolution. Using the line strength of the molecular transitions as well as the sensitivity and bandwidth of our instrument, we determine the noise equivalent concentration (NEC) at typical measurement settings of 0.014 cm$^{-1}$ unapodized resolution, 600 Torr of $N_2$ background pressure, and 30 s total integration time (5 averaged scans for reference and sample). The NEC for a small collection of different molecular species within our spectral range is summarized in Tab. 1 (column 3), which shows that ppb-level NEC limits are obtained for most molecules in only 30 s.

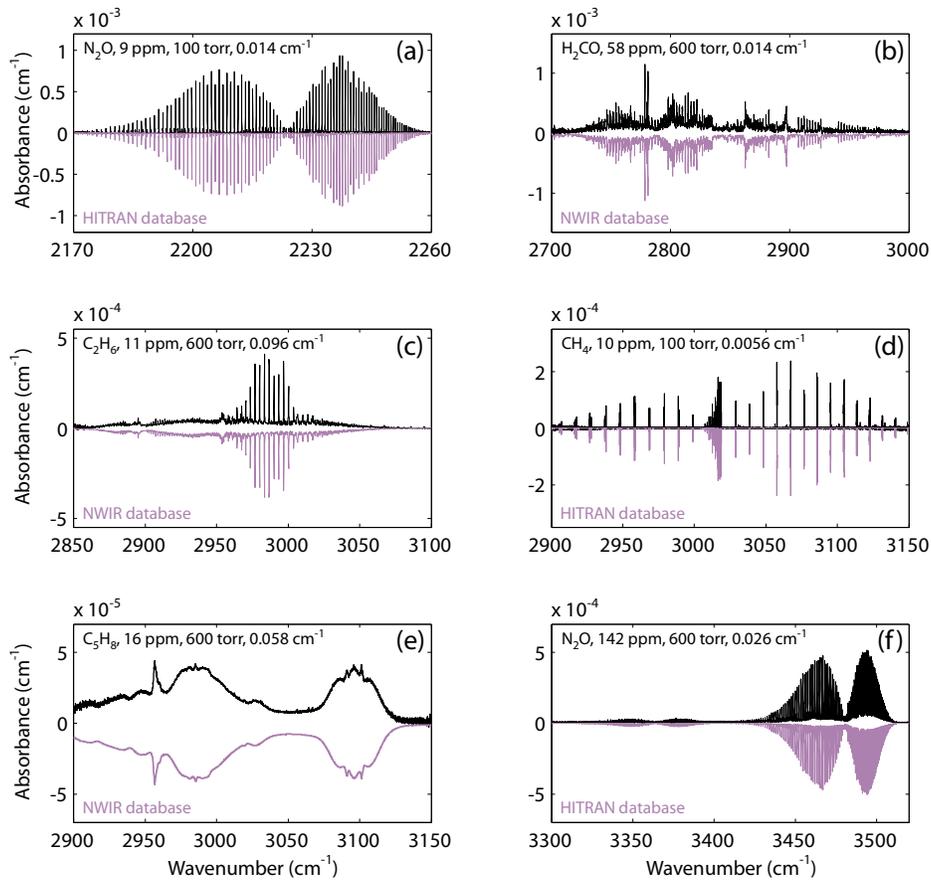

Fig. 2. Collection of absorption spectra throughout the operating range of our mid-IR frequency comb Fourier transform spectrometer (black) and comparison to HITRAN or NWIR-based fits (purple, plotted negative for clarity). The trace concentration, nitrogen pressure, and spectral resolution are designated in the panels; (a) $\nu_3$ band of nitrous oxide (N$_2$O) at a concentration of 9 ppm, a pressure of 100 Torr N$_2$, and a resolution of 0.014 cm$^{-1}$; (b) formaldehyde (H$_2$CO), 58 ppm, 600 Torr, 0.014 cm$^{-1}$; (c) ethane (C$_2$H$_6$), 11 ppm, 600 Torr, 0.096 cm$^{-1}$; (d) methane (CH$_4$), 10 ppm, 100 Torr, 0.0056 cm$^{-1}$; (e) isoprene (C$_5$H$_8$), 16 ppm, 600 Torr, 0.058 cm$^{-1}$; (f) $2\nu_2+\nu_3$ and $\nu_1+\nu_3$ band of N$_2$O, 142 ppm, 600 Torr, 0.026 cm$^{-1}$. All spectra are obtained with a single wavelength setting of the OPO except (a), which is a combination of two measurements at different central wavelengths.

The best unapodized resolution of our FC-FTS is currently 0.0056 cm$^{-1}$ (168 MHz), which is limited by the memory of the data acquisition card. In Fig. 3(a) we show a zoom of the methane Q-branch [from Fig. 2(d)] to emphasize the excellent resolving power (data in black, HITRAN-based fit in red). Fig. 3(b) displays a further magnified spectrum that illustrates the actual data points (open circles); the width of the absorption lines is limited mostly by pressure broadening (here, 100 Torr N$_2$). The optical delay length of our system allows us in principle to resolve single comb lines. Then, the advantage of the frequency comb in terms of resolution becomes apparent. Instead of scaling with the delay length, the resolution limit will be given by the linewidth of individual comb teeth [10,12,36], which is ~1×10$^{-6}$ cm$^{-1}$ (40 kHz) [31].

Table 1. Detection limits for a collection of important molecules within the spectral range of our mid-IR FC-FTS. The third column shows the noise equivalent concentration (NEC) results, which are based on the measured absorption sensitivity of $3.8\times10^{-8}$ cm$^{-1}$ Hz$^{-1/2}$ per spectral element under typical experimental conditions of 600 Torr N$_2$ pressure, 0.014 cm$^{-1}$ unapodized spectral resolution, and 30 s of total integration time (5 averages). The fourth column shows the theoretically estimated multi-line detection limits (see Appendix) using the same noise and resolution conditions as those for column three. The fifth column displays the experimentally obtained concentration detection limits from the multi-line fits to measured spectra. The disagreement between the theoretical and experimental values is due to the fact that some spectra were not measured at the best possible noise performance of the system. Note that significantly lower detection limits are achievable with longer integration times (tested to at least 6 min).

| Molecule | Band center (cm$^{-1}$) | NEC[a] (ppb) | Theor. multi-line DL[b] (ppb) | Exp. multi-line DL[b] (ppb) |
|---|---|---|---|---|
| CH$_4$ | 3020 | 22 | 2.0 | 5 |
| C$_2$H$_6$ | 2990 | 23 | 1.8 | 18 |
| C$_5$H$_8$ | 3000 | 370 | 4.3 | 7 |
| H$_2$CO | 2780 | 54 | 2.3 | 40 |
| CH$_3$OH | 2950 | 350 | 3.6 | 40 |
| C$_2$H$_5$OH[c] | 2970 | 250 | 2.8 | n/a |
| (CH$_3$)$_2$CO[c] | 2950 | 370 | 9.1 | n/a |
| N$_2$O | 2220 | 8 | 0.50 | 5 |
| CO$_2$ | 2350 | 2.2 | 0.20 | 0.50 |
| CO[c] | 2150 | 14 | 2.2 | n/a |
| NH$_3$[c] | 3380 | 400 | 25 | n/a |
| H$_2$O[c] | 3750 | 60 | 5.3 | n/a |

[a] NEC, noise equivalent concentration
[b] DL, detection limit
[c] molecules not measured in the experiment; detection limits projected from results of other species in the same spectral region

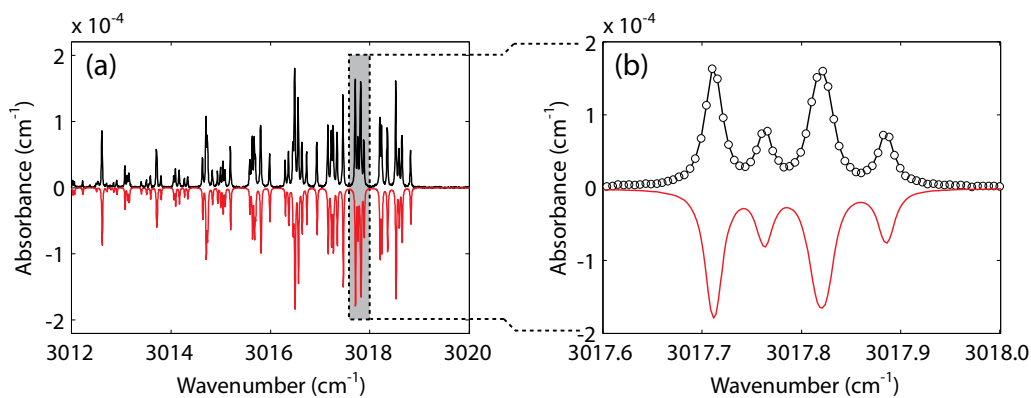

Fig. 3. (a) A zoom of the methane Q-branch spectrum at the system's highest resolution of 0.0056 cm$^{-1}$ (168 MHz) and an N$_2$ pressure of 100 Torr. (b) A further magnified view of one of the line features; open circles in black are measured single data points; fitted HITRAN-based spectrum is shown in red and plotted negative for clarity.

To test the repeatability of our instrument, we take ten independent measurements of a methane mixture certified as 10 ppm (± 5%) in $N_2$ at a pressure of 600 Torr, 0.014 cm$^{-1}$ resolution, and 30 s integration time and extract the concentration of each dataset from a numerical fit of a HITRAN reference spectrum. We determine the average $CH_4$ concentration to 10.107 ppm with a standard deviation of 0.019 ppm, which is in agreement with the specified concentration of the gas mixture and demonstrates that the precision of the entire system (optical and gas handling) is better than 0.2%.

### 3.3 Multi-line fitting advantage

An important and already recognized advantage of broadband spectroscopy is that the ability of detecting multiple absorption peaks over a wide region allows one to achieve a lower detectable concentration than with single-peak measurements [37-39]. A theoretical estimation of the resulting benefit can be found in the Appendix. In general, the multi-line advantage scales with the integrated area under the absorption curve. To confirm this finding, we take the uncertainty of the fitted concentrations of the ten measurements of $CH_4$, which are extracted by utilizing the entire recorded spectrum, and compare it against the uncertainty when using only one of the strong peaks of the P-branch. The single-peak fits give an uncertainty of 20-30 ppb, which is consistent with the NEC. In contrast, the multi-line fit yields an uncertainty of only 7 ppb, which is significantly lower than the NEC and confirms that analysis of the entire absorption spectrum can indeed decrease the detection threshold below the classical limit given by the single-line S/N ratio. As an example, Fig. 4(a) shows a spectrum of a diluted sample of $CH_4$ (275 ppb) which has a very low S/N ratio. Nevertheless, the multi-line fit can determine the sample concentration with an uncertainty of only 5 ppb (1.8% relative uncertainty). The inset demonstrates that this relative uncertainty is significantly less than the inverse of the S/N ratio on a single peak; the residuals of the fit are plotted in Fig. 4(b), showing a structureless background. The detection limits obtained from the multi-line fitting procedure for other measured molecules are summarized in the fifth column of Tab. 1. The fourth column in Tab. 1 shows the theoretical estimates for detection limits under the same conditions of noise and resolution used for the third column. The scaling of the detection limit indicated by the theory corresponds well with the experimental numbers: for example, the improvement in detection limit for $C_5H_8$ is 12 times as high as the one for $CH_4$ due to the much larger integrated absorption of the continuous spectrum of isoprene. The experimental results shown in the fifth column of the table differ from the theoretical estimates due to the fact that some spectra were not recorded during the best possible noise performance of the system, which is mostly due to varying temperature stability of the laboratory.

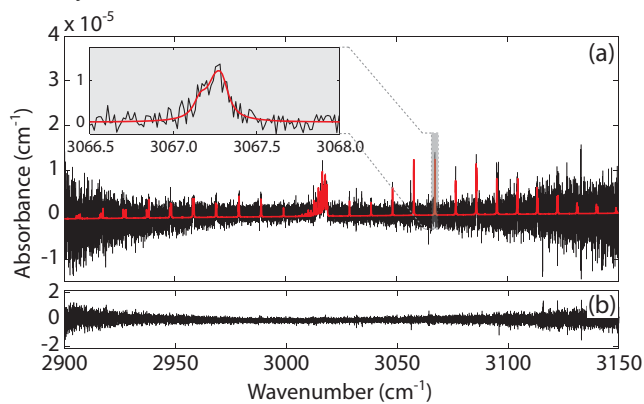

Fig. 4. (a) Measurement of $CH_4$ at a highly diluted concentration of 275 ppb. Despite the low S/N ratio, the detection and fitting of multiple lines in the absorption band allow a reliable determination of the concentration with only 5 ppb uncertainty. The inset shows a comparison of measured data (black) and fit from HITRAN reference (red) at one single feature. (b) The residual of the multiple-line fit.

*3.4 Determination of absolute concentrations of a gas mixture*

Most real-world samples measured for atmospheric science, breath analysis, or trace gas monitoring contain more than one species in a particular spectral window. The overlapping absorption features of different molecules make their unambiguous detection nearly impossible if sufficient spectral coverage and resolution are not available. The C-H-stretch region around 3000 cm$^{-1}$, for instance, contains absorption features of practically any hydrocarbon with sufficient abundance. To demonstrate the capability of our spectrometer, we measure the absorption spectrum of a mixture of trace amounts of formaldehyde, methanol, and water in nitrogen and determine their absolute concentrations. The spectrum was recorded at a resolution of 0.014 cm$^{-1}$ and an N$_2$ pressure of 600 Torr under a total integration time of 30 s. Here, methanol – as a molecule without a clear line spectrum – would represent a significant challenge for a cw laser measurement. Fig. 5(a) shows the measured spectrum of the H$_2$CO/CH$_3$OH/H$_2$O-mixture (black curve) in the spectral range from 2900–3050 cm$^{-1}$ and the overall fitted result (gray, plotted negative for clarity). Model spectra created from HITRAN (for H$_2$CO and H$_2$O) and NWIR (for CH$_3$OH) of the individual components at their fitted concentrations are shown in magenta (H$_2$CO), green (CH$_3$OH), and blue (H$_2$O), respectively. The overall fit is remarkably accurate, as demonstrated by the fit residuals plotted in Fig. 5(b). Although the spectrum contains many overlapping features and some of them (from CH$_3$OH) do not exhibit sharp lines, the broad bandwidth and high spectral resolution of our FC-FTS in combination with the multi-line fitting procedure enable precise determination of the concentrations of all three components. The obtained values are (47.1 ± 0.1) ppm for H$_2$CO, (58.30 ± 0.04) ppm for CH$_3$OH, and (813 ± 7) ppm for H$_2$O.

Another important feature of the current system is the large dynamic range in concentration detections. To demonstrate this, we measured a sample of laboratory air, which contains atmospheric methane; however, the high abundance of water vapor, whose absorption lines strongly overlap with the CH$_4$ band at 3020 cm$^{-1}$, imposes a particular challenge for trace gas detection. Although the strongest H$_2$O lines are already close to saturation, the fitting algorithm is able to extract both concentrations, in particular that of CH$_4$. Fig. 6(a) shows the recorded spectrum at full scale (black) and the HITRAN-based fits for water (blue) and methane (red); the magnified view in Fig. 6(b) reveals the weak CH$_4$ features in the spectrum. The concentrations obtained from the fit are (2.02 ± 0.01) ppm for CH$_4$ and (1.03 ± 0.05)% for H$_2$O.

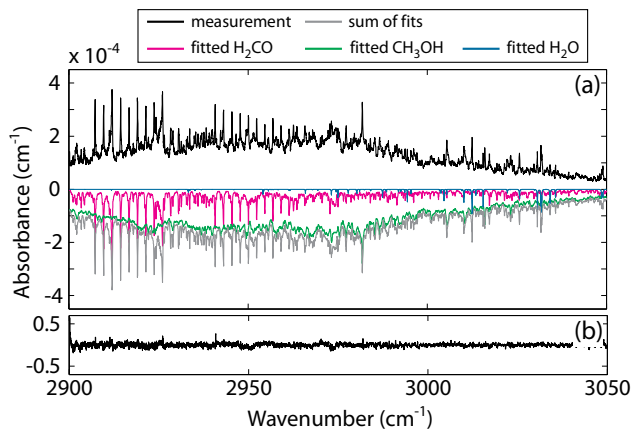

Fig. 5. (a) Measured spectrum of a mixture of formaldehyde, methanol, and water at an N$_2$ pressure of 600 Torr and a resolution of 0.014 cm$^{-1}$ (black), as well as the fitted spectra based on HITRAN and NWIR of the three components (H$_2$CO, magenta; CH$_3$OH, green; H$_2$O, blue) and their sum (gray). The fit gives abundances of (47.1 ± 0.1) ppm H$_2$CO, (58.30 ± 0.04) ppm CH$_3$OH, and (813 ± 7) ppm H$_2$O. (b) The residual of the overall fit displays no remaining structure.

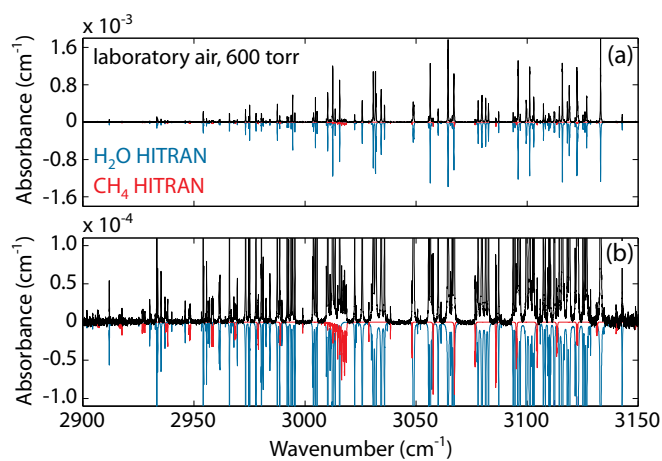

Fig. 6. Measurement of laboratory air at 600 Torr recorded with a resolution of 0.014 cm$^{-1}$ (black) containing 2.02 ppm of methane and 1.03% of water, as well as HITRAN-based fits of H$_2$O (blue) and CH$_4$ (red). (a) The full scale of absorbance displays mainly water lines, and the magnified scale in (b) reveals the weak methane lines among the much stronger water peaks.

## 4. Conclusions

With the present system, we advance high-resolution and high-sensitivity spectrometry with frequency combs for the first time into the important molecular fingerprint region. The mid-infrared FC-FTS is capable of detecting a wide variety of molecular species with sensitive and reproducible determination of their concentration in the spectral range between 2100 cm$^{-1}$ and 3700 cm$^{-1}$ with a maximum resolution of 0.0056 cm$^{-1}$. Importantly, we demonstrate trace detection of molecular species exhibiting both sharp spectral lines and continuous bands. In particular, the latter task is extremely difficult with cw lasers or near infrared comb systems. The simultaneous acquisition of multiple spectral lines over a wide spectral window (~50-400 cm$^{-1}$) has allowed us to extend the detection limit to below the actual noise floor by taking advantage of multi-line fitting to all present absorption lines or the full shape of the spectral feature. The absorption sensitivity of the instrument of $3.8 \times 10^{-8}$ cm$^{-1}$ Hz$^{-1/2}$ per detection element is sufficiently high to achieve low-ppb NECs for many molecules in 30 s total integration time (including reference and sample spectrum) at an unapodized resolution of 0.014 cm$^{-1}$. The significant decrease in acquisition time compared to standard FTIR is a direct result of the high spectral brightness and spatial coherence of the frequency comb source. These properties allow us to overcome the thermal noise of IR detectors and to achieve excellent throughput in the multi-pass cell and the interferometer. Most notably, the étendue of the FC-FTS does not decrease with optical resolution in contrast to classical FTIR [7] with incoherent light sources, which makes our system particularly attractive for high-resolution applications. Detection limits in the sub-ppb range are also possible by increasing the number of scans; we experimentally verified that the noise averages down with $\tau^{-1/2}$ for at least 6 minutes of integration time $\tau$. In this case, the multi-line detection limit for CO$_2$, for instance, drops to 0.14 ppb at a resolution of 0.014 cm$^{-1}$. Note that the data acquisition time can also be significantly shortened if an application does not require as high spectral resolution.

With the capability of providing almost real-time and broadband spectral data with high sensitivity in the important molecular fingerprint region, our mid-IR FC-FTS represents a practical system that is ready to replace FTIR for many scientific applications ranging from atmospheric science to breath analysis. The high power of the femtosecond laser source could also allow spectral broadening in photonic crystal fibers to create a spectroscopy tool that can span thousands of wavenumbers at a time [40,41]. Since the phase-locked mid-infrared OPO emits a stable frequency comb [31], the addition of a high-finesse femtosecond enhancement cavity for sample detection is feasible, which will further advance the sensitivity of the system [9-12,21,36,42]. Furthermore, broadband versions of Doppler-free methods such as saturation spectroscopy [42] or mid-infrared polarization spectroscopy [42-44] are future possibilities with this system. Therefore, this work provides an encouraging outlook for comb-based spectroscopy instruments to deliver future scientific results that are unattainable with conventional techniques.

**Acknowledgments**

We thank R. Fox for the loan of the Herriott cell and D. L. Osborn, M. J. Thorpe, N. R. Newbury, S. A. Diddams, and R. Ciuryło for valuable discussions. We further thank Matheson-Trigas for providing certified sample gas mixtures. F. A. was partially supported by the Alexander von Humboldt Foundation. P. M. acknowledges a fellowship from the Polish Ministry of Science and Higher Education. A. F. holds a fellowship from the Swedish Research Council. K. C. C. is supported by the National Science Foundation Graduate Fellowship. This project is funded by the Air Force Office for Scientific Research, the Defense Advanced Research Projects Agency, the Agilent Foundation, the National Institute for Standards and Technology, and the National Science Foundation.

**Appendix: Theoretical estimation of concentration detection limit for multi-line fit**

Fitting a model spectrum to the measurement data allows decreasing the uncertainty of the fitted concentration parameter compared to the experimental uncertainty of the single point given by the noise of the measured spectrum. Applied to single peaks – usually by means of a nonlinear least-squares method and Marquardt algorithm – it is a common way of analysis of spectroscopic data. When applied to extremely broadband spectra, however, the multi-line analysis can significantly decrease the concentration detection limits, as it is shown in Tab. 1.

To underline our experimental findings, we derive an equation that allows the estimation of the concentration detection limit obtained with multi-line analysis via the known structure of a molecule's absorption spectrum. We assume that $\alpha(v_i)$ is the absorption spectrum measured at $k$ discrete frequencies $v_i$ with a standard deviation $S_\alpha$, which is uniform over the spectrum. The model spectrum $\alpha_T(v_i)$ is calculated for a given sample concentration $C$ at the experimental pressure and temperature. For simplicity we assume that there is no baseline in our spectrum and that the model spectrum has exactly the same frequency axis as the measured spectrum. We would like to find the parameter $N$ (which will be used to estimate the concentration), for which differences between measured and model spectra multiplied by this parameter will be minimized. To do this, we fit the value of parameter $N$ using a linear least-squares method to minimize the function

$$F(N) = \sum_{i=1}^{k} \left[ \alpha(v_i) - N\alpha_T(v_i) \right]^2 = \min, \qquad (1)$$

$$\frac{dF}{dN} = \sum_{i=1}^{k} 2\left[ \alpha(v_i) - N\alpha_T(v_i) \right]\left[ -\alpha_T(v_i) \right] = 0. \qquad (2)$$

The fitted value of $N$ equals to

$$N = \frac{\sum_{i=1}^{k} \alpha(v_i)\, \alpha_T(v_i)}{\sum_{i=1}^{k} \alpha_T^{\,2}(v_i)}. \qquad (3)$$

We can estimate the standard deviation of the fitted parameter $N$ by

$$S_N^2 = \sum_{i=1}^{k} \left[ \frac{\partial N}{\partial \alpha(v_i)} \right]^2 S_\alpha^2. \qquad (4)$$

Since

$$\frac{\partial N}{\partial \alpha(v_i)} = \frac{\alpha_T(v_i)}{\sum_{i=1}^{k} \alpha_T^{\,2}(v_i)}, \qquad (5)$$

we obtain

$$S_N^2 = \frac{\sum_{i=1}^{k} \alpha_T(\nu_i)^2 S_\alpha^2}{\left[\sum_{i=1}^{k} \alpha_T^2(\nu_i)\right]^2} = \frac{S_\alpha^2}{\sum_{i=1}^{k} \alpha_T^2(\nu_i)} .$$ (6)

The standard deviation of the fitted parameter $N$ equals

$$S_N = \frac{S_\alpha}{\left[\sum_{i=1}^{k} \alpha_T^2(\nu_i)\right]^{1/2}} .$$ (7)

The value of the concentration obtained from the fit equals $N\,C$, whereas the concentration detection limit obtained from the multi-line fit is equal to $S_N\,C$. Despite the fact that Eq. (7) is strictly valid only for fitting one linear parameter, it is still a good estimate of the concentration detection limit in the case of multiple fitting parameters, provided the molecular absorption spectrum $\alpha_T(\nu_i)$ and a good estimate for the noise of the experimental setup $S_\alpha$ are known. It should be noted that the estimate for the concentration detection limit will be as good as the estimate of the standard deviation of noise $S_\alpha$. For example, if we look at the spectrum of methane [Fig. 2(d)] and calculate the standard deviation of the residuals of the fit, the experimental uncertainty of the concentration agrees within 0.02% with the theoretical estimate. In general, the residuals are not known a priori; however, we can estimate the standard deviation of the noise from a portion of the spectrum without absorption features.

We further note that according to Eq. (7), the multi-line analysis improvement of the detection limit depends on the shape of the spectrum and scales with its integrated absorption. This finding becomes apparent when comparing spectra of methane [Fig. 2(d)] and isoprene [Fig. 2(e)], which exhibit clearly different structure. The numbers from Tab. 1 show that the multi-line analysis improves the detection limit by factors of 11 for $CH_4$ and 93 for $C_5H_8$ owing to the much larger integrated area of the continuous isoprene absorption spectrum.